\documentstyle[stwol,epsf]{article}

\def\npb#1{Nucl.~Phys.~{\bf B#1}}

\def\prd#1{Phys.~Rev.~{\bf D#1}}

\def\e{\epsilon}

\def\fig#1{Fig.~\ref{#1}}

\def\e3{$\epsilon_3$}
\def\tb{tan$\beta$}
\def\bsgam{$b\rightarrow s\gamma$ }
\def\ch2{$\chi^2$}
\def\Bk{$\hat B_K$}
\def\co#1{{\ifmmode{\cal O}_{#1}\else${\cal O}_{#1}$\fi}}
\def\mx{\mbox}

\newdimen\unit
\def\point#1 #2 #3{\vbox to0pt{\kern-#2\unit
  \hbox{\kern#1\unit#3}\vss}
 \nointerlineskip}

\newcommand{\be}{\begin{equation}}
\newcommand{\ee}{\end{equation}}
\newcommand{\bea}{\begin{eqnarray}}
\newcommand{\eea}{\end{eqnarray}}

\begin{document}

\begin{titlepage}

\hfill OHSTPY-HEP-T-96-037 \\ 
\mbox{ } \hfill November 1996 \\

\vspace{2.0 cm}

\begin{center}

\renewcommand{\thefootnote}{\fnsymbol{footnote}}

{\Large\bf Supersymmetric Grand Unified Theories and 
               Global Fits to Low Energy Data}

\vspace{1.0 cm}

{\large\bf  T. Bla\v{z}ek~\footnote
            {On leave of absence from the Dept.\ of Theoretical Physics, 
            Comenius Univ., Bratislava, Slovakia; current e-mail address: 
            {\em blazek@mps.ohio-state.edu}} and S. Raby}\\ 
\bigskip
{\em Department of Physics, The Ohio State University, 174 W. 18th Ave., Columbus, OH 43210}

\vspace{2,3 cm}

{\bf Abstract}

\end{center}

We present a self-consistent $\chi^2$ analysis of several supersymmetric 
(SUSY) grand unified theories recently discussed in the literature. We 
obtain global fits to low energy data, including gauge couplings, fermion 
masses and mixing angles, gauge boson masses and $BR(b\rightarrow s\gamma)$. 
One of the models studied provides an excellent fit to the low energy data 
with $\chi^2\sim 1$ for 3 degrees of freedom, in a large region of the 
experimentally allowed SUSY parameter space. We also discuss 
the consequences of our work for a general MSSM analysis at the $Z$ scale.

\vfill

\setcounter{footnote}{0}
\renewcommand{\thefootnote}{\arabic{footnote}}

\end{titlepage}

%
%

\title{Supersymmetric Grand Unified Theories and 
               Global Fits to Low Energy Data}

\author{T. Bla\v{z}ek and  S. Raby}

\address{Department of Physics, The Ohio State University, 174 W. 18th Ave., Columbus, OH 43210}


\twocolumn[\maketitle\abstracts{
We present a self-consistent $\chi^2$ analysis of several supersymmetric 
(SUSY) grand unified theories recently discussed in the literature. We 
obtain global fits to low energy data, including gauge couplings, fermion 
masses and mixing angles, gauge boson masses and $BR(b\rightarrow s\gamma)$. 
One of the models studied provides an excellent fit to the low energy data 
with $\chi^2\sim 1$ for 3 degrees of freedom, in a large region of the 
experimentally allowed SUSY parameter space. We also discuss 
the consequences of our work for a general MSSM analysis at the $Z$ scale. 
}]

\section{Introduction}

There is no doubt that the Standard Model (SM) describes physical 
processes at the highest available energies with very good precision. Yet it 
contains some 18 independent parameters which remain completely 
undetermined and await derivation from a more fundamental theory. 
13 of these parameters are related to the fermion mass sector which is  
clearly distinguished by an amazingly simple regularity in the hierarchy 
of masses and mixing angles of the three fermionic families.

This observed pattern may provide a clue to a more fundamental 
theory at some higher scale $M$ ($M\sim M_{Planck}$ or $M_{string}$ or $M_{GUT}$). 
Our main hypothesis is that only a {\em small} set of effective mass operators below $M$
dominates in the Yukawa matrices, subsequently leading to the fermion mass 
matrices at the scale of electroweak symmetry breaking. These effective operators 
are of utmost interest.  They can lead to the 
reconstruction of the full effective field theory at the scale $M$ and 
provide the  matching conditions for a fundamental string theory, believed to 
be the ultimate quantum theory incorporating gravity together 
with the gauge interactions of quarks and leptons. 

SO(10) supersymmetric theories are excellent candidates 
for such a theory below the string scale. They maintain the successful
prediction for gauge coupling unification and provide
a powerful framework for predictive theories of fermion
masses and mixing angles. This is because all the fermions
of a single family are contained in the 16 dimensional
representation of SO(10) - thus fermion mass matrices are
related by symmetry. In the most predictive theories,
the ratio of Higgs vevs - tan$\beta$ - is large, and the
top quark is naturally heavy as found experimentally. However, 
as a consequence of large tan$\beta$ there are potentially large
supersymmetric weak scale threshold corrections which could
 play an important role in fitting the fermion masses and
mixings. Thus a self-consistent analysis necessarily includes
the dimensionful soft SUSY breaking parameters, in addition to the 
dimensionless gauge and Yukawa couplings.

\section{Connecting GUT Scale Physics with Low Energy Observables}

Here we present the results of such a complete top-bottom analysis.
It starts at the GUT scale $M_G$, which is a free parameter itself, with 
unified gauge coupling $\alpha_G$, $n_y$ free parameters entering 
the Yukawa matrices 
\footnote{Clearly, $n_y$ is model dependent.}, 
                    and with five universal soft SUSY breaking parameters 
$\mu,\: m_0,\: M_{1/2},\: A_0$ and $B\mu$. In addition, we introduce 
\e3 as a one loop GUT threshold correction to $\alpha_s(M_G)$, 
    \footnote{By definition, $M_G$ is to be understood as the scale 
              where $\alpha_1$ and $\alpha_2$ 
              are the same, and equal to $\alpha_G$ .}
and non-universal Higgs masses $m_{H_d}$ and $m_{H_u}$ .
The dimensionless (dimensionful) couplings are run down to the $Z$ 
scale using two (one) loop renormalization group equations (RGEs) of MSSM. 
We have checked at a few selected points that the corrections to our results 
obtained by using two loop RGEs for dimensionful couplings are, in fact, 
insignificant. At the $Z$ scale we match the MSSM directly \cite{cpp} to the 
$SU(3)_c\times U(1)_{EM}$, thus leaving out the SM as an effective theory 
 on our way down to the experimentally measured low energy data listed 
in table \ref{t:obs}. 
\protect
\begin{table}[tb]
\caption{Experimental values of the twenty low energy observables 
         entering the \ch2 function, together with the corresponding 
         standard deviations.}
\label{t:obs}
\begin{tabular}{|c|c|c|}
\hline
Observable & Central value & $\sigma$ \\
 \hline
$M_Z $            &  91.186       & 0.46     \\
$M_W $            &  80.356       & 0.40     \\
$G_{\mu} $        &   $0.11664 \times 10^{-4}$  &  $0.0012 \times 10^{-4}$ \\
$\alpha_{EM}^{-1}$  &  137.04       & 0.69     \\
$\alpha_s(M_Z)$   &  0.118        & 0.005    \\
$\rho_{new}$      & $-0.6 \times 10^{-3}$  &  $2.6 \times 10^{-3} $    \\
\hline
$M_t  $           &  175.         & 6.      \\
$m_b(M_b)  $      &    4.26       & 0.11     \\
$M_b - M_c $      &    3.4        & 0.2      \\
$m_s $            &  180.         &  50.      \\
$m_d/m_s$         &  0.05         &  0.015    \\
$Q^{-2} $         &  0.00203      &  0.00020     \\
$M_{\tau}$        &  1.777        &   0.0089   \\
$M_{\mu} $        & 105.66        &   0.53    \\
$M_e  $           &  0.5110       &   0.0026   \\
$V_{us}$          &  0.2205       &  0.0026     \\
$V_{cb}$          &  0.0392       &  0.003     \\
$V_{ub}/V_{cb}$   &  0.08         &   0.02    \\
\Bk               &  0.8          &    0.1     \\
\hline
BR(\bsgam)        &  $0.232\times 10^{-3}$ &  $0.092\times 10^{-3}$ \\ 
\hline
\end{tabular}
\end{table}
At the $Z$ scale we use the tree level conditions for 
electroweak symmetry breaking to fix $v$ and \tb . Then, still within 
the MSSM, we calculate the amplitude for the process \bsgam, one loop 
corrected $W$ and $Z$ masses, corrections to the $\rho$ parameter from 
new physics outside the SM, and one loop corrected $G_{\mu}$, where we 
leave out the SUSY vertex and box contributions to the $\Delta r$ 
parameter. When crossing the $Z$ scale, we compute 
the complete one loop threshold corrections to  $\alpha_s$ and 
$\alpha_{EM}$, whereas only those one loop threshold corrections to the 
fermion masses and mixings enhanced by \tb\ are computed. 
\cite{bpr} Masses and couplings which require 
further running are renormalized down to their apropriate scales using 
three loop QCD and one loop QED RGEs. The branching ratio \mbox{BR(\bsgam)} is 
renormalized down to the scale $M_b$ using the leading log approximation. 
Note that some 
observables entering the \ch2 function (see table \ref{t:obs}) are known 
so well that we have to assign a theoretical, instead of experimental, 
error as their standard deviation. This is true for $M_Z,\: M_W,\: 
\alpha_{EM},\: G_{\mu}$ and the charged lepton masses. We estimated 
conservatively our theoretical error to be 0.5\% based on the uncertainties 
from higher order perturbation theory and from the performance of our 
numerical analysis. The error on $G_{\mu}$ is estimated to be within 1\% 
due to the fact that in addition to the uncertainties mentioned above 
we neglect SUSY vertex and box corrections to $\Delta r$.
Also note that $\epsilon_K$, the observable of $CP$ violation, has been 
replaced by a less precisely known hadronic matrix element \Bk . 
Thus our theoretical value of \Bk is defined as that value needed to agree 
with $\epsilon_K$ for a set of fermion masses and mixing angles derived 
from the GUT scale. Finally note, that the light quark masses are replaced by 
their ratios (at the scale 1GeV),  and $m_c(M_c)$ by the difference 
$M_b-M_c$, since the latter quantities are known to better accuracy. 
$Q^{-2}=(m_d^2-m_u^2)/m_s^2$ is the Kaplan-Manohar-Leutwyler ellipse 
parameter.

In addition, the \ch2 function is increased significantly by a special penalty 
whenever a sparticle mass goes below its current experimental limit.

\section{Model 4 of ADHRS}

\begin{figure}[tb]
\setlength{\epsfxsize}{90mm}\epsfbox[80 150 470 595]{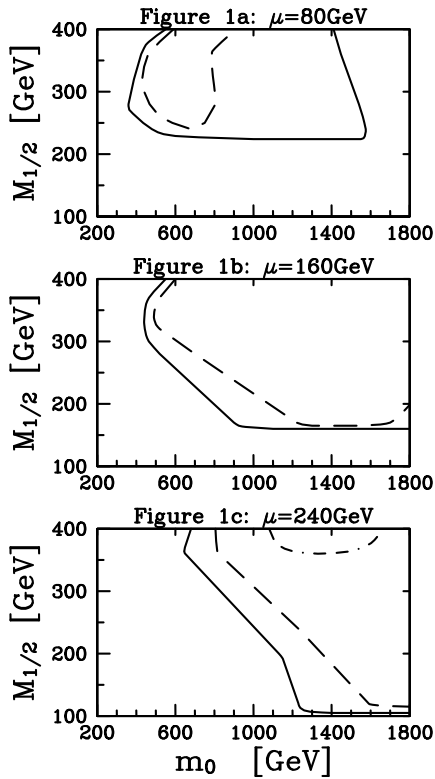}
\caption{Model 4 global analysis results for different values of $\mu(M_Z)$.
         Solid (dashed, double-dash-dotted) lines represent contour lines of
         constant \ch2 = 15 ( 14, 13 ) / 5dof.}
\label{fH4_}
\end{figure}

We have analyzed several models of fermion masses. First, we have studied 
the best working model of ADHRS \cite{adhrs}, model 4. The model is defined 
by four effective operators at the GUT scale ($A$'s stand for adjoint states 
and $S$ for a singlet state):  \\
\noindent
$\co{33}=16_3\:10\:16_3,\hfill 
 \co{23}=16_2\:{A_2\over \tilde A}\:10\:{A_1\over \tilde A}\:16_3,\\ 
\noindent
 \co{12}=16_1\:({\tilde A \over S})^3\:10\:({\tilde A \over S})^3\:16_2 
\;\; \;$ and by 
one of six possible \co{22} operators (they all give the same 
0:1:3 Clebsch relation between up quarks, down quarks and charged
leptons responsible for the Georgi-Jarlskog relation).
Each operator enters with its own complex coefficient, and we can rotate 
away three independent phases. Hence in model 4 $\; n_y$=5; the number of initial 
parameters for the \ch2 analysis is 15 and we are left with 5 degrees of 
freedom. \fig{fH4_} shows the contour lines of constant \ch2 in the 
$m_0\,-\,M_{1/2}$ plane for different values of $\mu$. 
No substantial improvement of the performance of model 4 can be achieved 
by neglecting one out of the twenty low energy data given in table 
\ref{t:obs}.~\cite{brcw} On the other hand we found that a significant 
improvement is possible by adding one new operator, contributing to the 
13 and 31 elements of the Yukawa matrices.

\section{Model 4c}

\begin{figure}[tb]
\setlength{\epsfxsize}{90mm}\epsfbox[80 150 470 595]{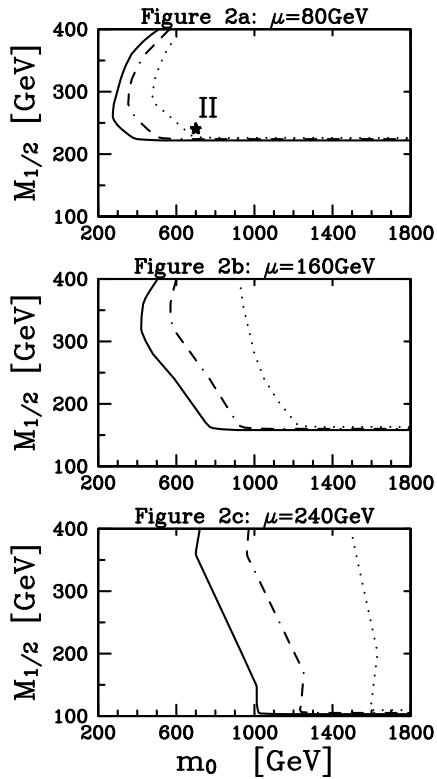}
\caption{Model 4c global analysis results for different values of $\mu(M_Z)$.
         Solid (double-dash-dotted, dotted) lines represent contour lines of
         constant \ch2 = 6 ( 3, 1 ) / 3dof.}
\label{fH4c}
\end{figure}

Next, we analyzed two models derived from the complete SO(10) SUSY GUTs 
discussed recently by Lucas and Raby\cite{lr}. The models were constructed 
as simple extensions of model 4. Different label 
\mbox{(a,b,...f)} refers to the different possible 22 operators.\footnote{Note, 
models d, e and f have the second family $16_2$ coupled directly to $10_1$ and 
a heavy $16$.  If this coupling is as large as the third generation Yukawa 
coupling, then we would obtain excessively large flavor changing neutral current 
processes, such as $\mu \rightarrow e \gamma$.  Thus these models were not 
considered in ref. [5]. }    In
the extension to a complete GUT the different 22 operators lead to 
inequivalent theories due to different U(1) charge assignments. 
When one demands ``naturalness'', i.e. includes all terms in the superpotential
 consistent with the symmetries of the theory one finds only one new
operator ($\co{13}$) for models 4a and 4c.  Model 4b, on the other hand, has 
no new operator and thus is equivalent to model 4 (above).  The 22 and 13 operators of
model 4c are 
\mbox{$ \co{22}=16_2\:{\tilde A \over S}\:10\:{A_1\over S}\:16_2$} and 
\mbox{$ \co{13}=16_1\:({\tilde A \over S})^3\:10\:{A_2 \over S}\:16_3\;\;$}. 
With the 13 operator $n_y$=7, which implies 3 degrees of freedom. The 
results of the global analysis are given in \fig{fH4c}, and in 
table \ref{t:m4cII} for a selected point II marked on \fig{fH4c}a. Model 4a 
is defined by different 22 and 13 operators and gives the fits 
with the best \ch2$\simeq$4-6 in most of the SUSY parameter space. 
(It yields also \ch2$\simeq$3 , but only for the corner in the SUSY 
parameter space with large $m_0,\: M_{1/2}$ and $\mu$.) \cite{brcw}

\section{Discussion and Conclusions}

The results of our analysis, as well as the whole project presented here, 
may be understood from two different perspectives. 

The emphasis of this work has been put on the study of the origin of fermion 
mass matrices, within the context of minimal SO(10) SUSY GUTs. 
The best working model of ADHRS, with just four effective operators in the 
Yukawa sector,can be assigned a confidence level of only 
3-4\%, in spite of the fact that each observable is within 
2$\sigma$ of its experimental value. The addition of a 13 operator may 
improve the performance of the model. Substantial improvement however, is 
not automatic, as was mentioned in the example of model 4a.
Nevertheless, we showed that model 4c provides an excellent fit to all 20 
observables of table \ref{t:obs}, with the confidence level better than 68\%
in a large region of the experimentally allowed SUSY parameter space. 
Note that the best fits extend to the region with very large $m_0$ where the effect 
of the SUSY threshold corrections to fermion masses and mixings is suppressed
by large squark and slepton masses. As a result, in this region the effective 
number of degrees of freedom in the fermionic sector is actually larger than 3, 
since there are 7 parameters ($A,B,C,D,\delta,E,\Phi$) in the Yukawa matrices 
determining the 13 low energy masses and mixings of the fermions. This means 
that the Yukawa 
sector of model 4c does actually a much better job than appears at first glance.  
Whether or not this particular model is close to the path Nature has 
chosen remains to be seen. 
\protect
\begin{table}[tb]
\caption{Model 4c - Results at Point II. \protect \mbox{                          } 
  Initial parameters:  \mbox{                           }
  1/$\alpha_G$ = 24.36, $M_G$ = 3.17$\cdot$10$^{16}$GeV, \e3 = -4.89\% ,
  A = 0.807, B = 5.44$\cdot$10$^{-2}$, C = 1.15$\cdot$10$^{-4}$, 
D = 4.94$\cdot$10$^{-4}$, $\delta$ = 5.71,
E = 1.31$\cdot$10$^{-2}$, $\Phi$ =  1.04, \ \ 
$\mu$ = 80GeV, $m_0$ = 700GeV, $M_{1/2}$ = 240GeV,\ \  
$m_{H_d}/m_0$ = 1.42, $m_{H_u}/m_0$ = 1.24,\ \  $A_0$ = 458.35GeV,\ \ $B\mu$ = 120.66GeV$^2$. 
The last column displays SUSY threshold corrections in \%. } 
\label{t:m4cII}
\begin{tabular}{|c|c|c|r|}
\hline
Observable & Computed & Partial \ch2 & S.t.c. \\
\hline
$M_Z$              &  91.12       & 0.02             &      \\
$M_W$              &  80.38       & \mx{  }$<$0.02   &      \\
$G_{\mu}$   &  1.166$\cdot$10$^{-5}$ &\mx{  }$<$0.02 &   \\
$\alpha_{EM}^{-1}$ &  137.0       & \mx{  }$<$0.02   &  1.43  \\
$\alpha_s(M_Z)$    &  0.1151      & 0.34             & 12.78  \\
$\rho_{new}$ &+1.87$\cdot$10$^{-4}$  & 0.09             &   \\
\hline
$M_t$              &  175.7       & \mx{  }$<$0.02   &  0.74  \\
$m_b(M_b)$         &    4.287     & 0.06             &  5.43  \\
$M_b - M_c$        &    3.440     & 0.04             &  7.56  \\
$m_s$              &  189.0       & 0.03             &  3.68  \\
$m_d/m_s$          &  0.0502      & \mx{  }$<$0.02   &  0.00  \\
$Q^{-2}$           &  0.00204     & \mx{  }$<$0.02   &  1.78  \\
$M_{\tau}$         &  1.776       & \mx{  }$<$0.02   & -2.08  \\
$M_{\mu}$          & 105.7        & \mx{  }$<$0.02   & -1.50  \\
$M_e$              &  0.5110      & \mx{  }$<$0.02   & -1.50  \\
$V_{us}$           &  0.2205      & \mx{  }$<$0.02   &  0.00  \\
$V_{cb}$           &  0.0400      & 0.07             &  1.58  \\
$V_{ub}/V_{cb}$    &  0.0772      & \mx{  }$<$0.02   &  0.00  \\
$\hat B_K$         &  0.8140      & \mx{  }$<$0.02   & -3.18  \\
\hline
BR(\bsgam) &  2.382$\cdot$10$^{-4}$ & \mx{  }$<$0.02 & \mbox{} \\
\hline
 \multicolumn{2}{|l} {\mx{ } TOTAL\ \ \ch2}
&\multicolumn{2}{l|} {\mx{ }             0.7306}   \\
\hline
\end{tabular}
\end{table}
One important test will be via the CP violating 
decays of the B meson. Model 4c predicts a value for sin2$\alpha\simeq0.95$ which 
is insensitive to the SUSY breaking parameters\cite{brcw}, whereas in the 
SM the value of sin2$\alpha$ is unrestricted\cite{ali}. Another important 
test may come from nucleon decay rates \cite{lr2}.

When one disregards the origin of the 
Yukawa matrices at the GUT scale, this work can also be viewed as an 
MSSM global fit in the large \tb\  regime. From this perspective, a 
special feature of our approach is that we are 
getting complete 3$\times$3 fermion and sfermion mass matrices at the 
SUSY breaking scale, instead of just the leading 33 elements. Viewed 
as MSSM global fits, our results suggest that there is no narrow, 
strongly preferred region in the SUSY parameter space. Hence, at present
one cannot make strong conclusions (with \tb\  large) about the masses 
of sfermions, which leaves open many channels for Tevatron and 
LEPII experiments.
\cite{brcw} It is also interesting to note that the best fits at a number of points 
give a low pseudoscalar Higgs mass\cite{brcw} and low values\cite{brcw} 
of $\alpha_s(M_Z)$ leaving the door open for a natural explanation of the 
1-2$\sigma$ increase in the partial width $Z\rightarrow b\bar b$, without any 
tuning of the initial parameters to obtain this effect.
Finally, by keeping the complete 3 $\times$ 3 mass matrices for both fermions and 
sfermions one can study flavor dependent processes {\em in a theory which fits the low 
energy data}; for example, rare $B$ and $K$ decays, $B - \bar B$ mixing, lepton 
flavor violating processes or even $Z \rightarrow b \bar b$.

At the present time, when direct evidence for physics beyond the SM 
evades experimental observations, global analysis serves as the best
test of new physics. Moreover, our results show that minimal SUSY SO(10) models 
remain among the best candidates for the theory of nature beyond the 
Standard Model.

\begin{flushleft}
{\bf Acknowledgements}
\end{flushleft}

The authors wish to thank Marcela Carena and Carlos E.M. Wagner for their contributions to this work.  This work was partially supported under DOE contract DOE/ER/01545-704.

\end{document}